# Proposal new area of study by connecting between information theory and Weber-Fechner law

Author name:  HaengJin Choe

Affiliation:  College of Engineering, University of Korea, Anamdong 5-ga, Seongbuk-gu, Seoul (Seoul 136-701 Korea)

Corresponding author:  HaengJin Choe. kchongup@korea.ac.kr

telephone number:  82-10-8958-0465

complete postal address:  College of Engineering, University of Korea, Anamdong 5-ga, Seongbuk-gu, Seoul (Seoul 136-701 Korea)

**Abstract:  Rough speaking, information theory deals with data transmitted over a channel such as the internet. Modern information theory is generally considered to have been founded in 1948 by Shannon in his seminal paper, "A mathematical theory of communication." Shannon's formulation of information theory was an immediate success with communications engineers. Shannon defined mathematically the amount of information transmitted over a channel [13]. The amount of information doesn't mean the number of symbols of data. It depends on occurrence probabilities of symbols of the data. Meanwhile, psychophysics is the study of quantitative relations between psychological events and physical events or, more specifically, between sensations and the stimuli that produce them [3]. It seems that Shannon's information theory bears no relation to psychophysics established by German scientist and philosopher Fechner. Here I show that**

**to our astonishment it is possible to combine two fields. And therefore we come to be capable of measuring mathematically perceptions of the physical stimuli applicable to the Weber-Fechner law. I will define the concept of new entropy. And as a consequence of this, new field will begin life.**

Keywords: information theory; Weber-Fechner law; **perceptional information theory**, **HaengJin entropy**

1  Introduction: The German physiologist Weber in 1834 carried out an exciting experiment [16]. Weber gradually increased the weight that the blindfolded man was holding. With doing so, Weber asked the subject when he first felt the increase of weight. Through this experiment, Weber found out that the least perceptible difference in weight was proportional to the starting value of the weight. For example, if a certain person barely feels the difference between 100g and 118g, then he barely feels the difference between 150g and 177g. In this case, $100:118 = 150:177$.

2.1  Material and methods: Fechner applied the result of the experiment of Weber to the measurement of sensation [5,14,15]. The Weber-Fechner law established in the field of psychophysics [3] attempts to describe the relationship between the physical magnitude of a stimulus and the perceived intensity of the stimulus. This fascinating law quantifies the human response to physical stimuli, which are weight, sound intensity, pitch, etc. As an example, the response of the human ear to changes in sound intensity is logarithmic. So sound intensity may be expressed in bels above the standard threshold of hearing intensity, in which case the standard threshold of hearing intensity is $10^{-12}$ W/m$^2$ [1,6,9,11]. We must remember that the bel scale of sound intensity is dimensionless and that the bel scale of sound intensity is logarithmic scale. As a second example, the response of the human ear to changes in pitch is logarithmic. In the musical instrument such as the piano, the frequencies of all notes form a geometric sequence so that musical intervals of every pair of successive notes are identical [1,6,9,11]. Free transposition is allowed because of such an equal temperament which is a musical temperament. In equal temperament every pair of adjacent notes has an identical frequency ratio, in which case the unit of frequency is Hz. As a third example, reaction of the eye to brightness of light is logarithmic. The stellar magnitudes $m_1$ and $m_2$ for two stars are related to the corresponding brightnesses $b_1$ and $b_2$ through the equation

$$m_2 - m_1 = \sqrt[5]{100} \log(b_1/b_2),$$

which results from the Pogson's proposal [10] that a difference of five magnitudes should be exactly defined as a brightness ratio of 100 to 1 [2,4,7,12,17]. We must remember that the stellar magnitude is dimensionless. Meantime, according to Shannon's information theory [13] we came to be able to quantify the amount of information transmitted over a channel. The ultimate goal of this paper is to append one more stimulus to the existing various stimuli applied to the Weber-Fechner law so that the Weber-Fechner law can be digitized. Digitized Weber-Fechner law, or more concretely, a combination of information theory and the Weber-Fechner law will lead us to the new field of study. But this additional stimulus and the corresponding perception are extremely extraordinary.

2.2 Material and methods: The Weber-Fechner law is derived as follows [8]. In Weber's experiment, let $S$ be weight at some instant and let $dS$ be the differential increment in weight. Let $dR$ be the differential change in perception. And let $k$ be a appropriate proportional constant, in which case $k$ is determined experimentally. The equation to express the result of the experiment of Weber is

$$dR = k(dS/S),$$

which is a differential equation. Integrating this equation yields

$$R = k \ln S + C,$$

in which case $C$ is a integration constant. Suppose that $S_0$ is the threshold of the stimulus below which a person doesn't perceive anything. Because $R = 0$ whenever $S = S_0$, $C$ must be equal to $-k \ln S_0$. Therefore, we can finally obtain

$$R = k \ln(S/S_0) = K \log_2(S/S_0),$$

which is the Weber-Fechner law expressed either physically or mathematically. Let us note that $R$ is a response to $S$ in this Weber-Fechner equation.

3  Theory: Shannon defined the amount of information of the symbol of occurrence probability $P$ as

$$I = -\log_2 P,$$

where the unit of $I$ is bits. But we can describe the above equation as

$$I = -\log_2(P/P_0),$$

where as a matter of course $P_0 = 1$. In the fourth equation, for all practical purposes $S \geq S_0$. In the sixth equation, for all practical purposes $P \leq P_0$. We may interpret $P_0$ as the threshold of $P$. Most importantly, we can interpret $I$ as a response to $P$ which is a special kind of stimulus. However this special stimulus differs in essence from ordinary physical stimuli in that it has no unit. Very important another fact is that the higher $P$ is, the lower $I$ is while the higher $S$ is, the higher $R$ is. But it is not essential that $I$ is a strictly decreasing function with respect to $P$ while $R$ is a strictly increasing function with respect to $S$. The essentially important facts are that $R$ and $I$ are logarithmic and that they both have the threshold values.

4  Conclusions: From these several viewpoints, consequently, I cannot choose but say that occurrence probability is a special kind of stimulus. It would be appropriate that we call occurrence probability mathematical stimulus in that occurrence probability has no physical unit. This mathematical stimulus is under the government of the Weber-Fechner law. By widening a range of applications of the Weber-Fechner law, we can combine information theory with the Weber-Fechner law of psychophysics. I named this new field **perceptional information theory**. Now we set $K$ to 1. Then we can say that when $S = S_0$, the amount of perception is 0bits and that when $S = 2S_0$, the amount of perception is 1bits. If the physical stimulus is weight, the problem is simple. From hence let's consider sound wave. Because sound wave is characterized by frequency and amplitude, considering only either frequency or amplitude may be meaningless. Thus, we must consider both simultaneously. For the sake of argument let's consider sound wave which has frequency $f$ and amplitude $a$. For this sound the total amount of perceptions is defiend as

$$\log_2(f/f_0) + \log_2(a/a_0) = \log_2(fa/f_0 a_0),$$

where the unit is bits. And the average amount of perceptions, which I named **HaengJin entropy**, is defined as

$$(\log_2(fa/f_0 a_0))/2,$$

where the unit is $\text{bits/response}$. **HaengJin entropy** is an arithmetic mean. While physics entropy and Shannon entropy are constrained by the fact that the total sum of probabilities is 1, in the above equation the physical quantities arbitrarily freely change. So it is impossible that we define **HaengJin entropy** with the form of Shannon entropy. That is, we cann't define **HaengJin entropy** as

$$(f_0/f)\log_2(f/f_0) + (a_0/a)\log_2(a/a_0).$$

This fact says that **HaengJin entropy** doesn't correspond completely to Shannon entropy. Nonetheless, it is useful to define the concept of **HaengJin entropy** since **HaengJin entropy** is linked directly with physics energy. We are easily able to understand such a fact. As an example, consider two sound waves. One wave has frequency $f_1$ and amplitude $a_1$, the other has frequency $f_2$ and amplitude $a_2$. One wave has **HaengJin entropy** $(\log_2(f_1 a_1/f_0 a_0))/2$, the other $(\log_2(f_2 a_2/f_0 a_0))/2$. So if $f_1 a_1 = f_2 a_2$, the two waves have the identical **HaengJin entropy**. But because $f_1 a_1 = f_2 a_2$ implies $f_1^2 a_1^2 = f_2^2 a_2^2$, $f_1 a_1 = f_2 a_2$ also means that the two waves have the identical energy. We can conclude from the above discussion that **HaengJin entropy** means the amount of reaction of a human being to energy of external physical stimuli and that the unit of **HaengJin entropy** is $\text{bits/response}$.

**Acknowledgement:** I would like to acknowledge to professor Heo who offered his opinion on this paper.

8[14]  S. S. Stevens, To honor Fechner and repeal his law, Science 133 (1961) 80–86.

[15]  J. Ward, An attempt to interpret Fechner's law, Mind 1 (1876) 452–466.

[16]  E. H. Weber, Der Tastsinn und das Gemeingefühl, 1851.

[17]  A. T. Young, How we perceive star brightnesses, Sky Telescope 79 (1990) 311–313.
**Vitae:**  I graduated in physics at Korea university. I got a master's degree in physics from Korea university.